# High-order chromatin architecture determines the landscape of chromosomal alterations in cancer

9/6/11

Geoff Fudenberg[1], Gad Getz[2], Matthew Meyerson[2,3,4,5], Leonid Mirny[6,7]

Author Affiliations:
[1] Harvard University, Program in Biophysics, Boston, Massachusetts.
[2] The Broad Institute of MIT and Harvard, Cambridge, MA 02142, USA.
[3] Harvard Medical School, Boston, MA 02115, USA.
[4] Department of Medical Oncology, Dana-Farber Cancer Institute, Boston, MA 02115, USA.
[5] Center for Cancer Genome Discovery, Dana-Farber Cancer Institute, Boston, MA 02115, USA.
[6] Harvard-Massachusetts Institute of Technology, Division of Health Sciences and Technology.
[7] Department of Physics, Massachusetts Institute of Technology, Cambridge, MA, USA

**The rapid growth of cancer genome structural information provides an opportunity for a better understanding of the mutational mechanisms of genomic alterations in cancer and the forces of selection that act upon them. Here we test the evidence for two major forces, spatial chromosome structure and purifying (or negative) selection, that shape the landscape of somatic copy-number alterations (SCNAs) in cancer[1]. Using a maximum likelihood framework we compare SCNA maps and three-dimensional genome architecture as determined by genome-wide chromosome conformation capture (HiC) and described by the proposed fractal-globule (FG) model[2]. This analysis provides evidence that the distribution of chromosomal alterations in cancer is spatially related to three-dimensional genomic architecture and additionally suggests that purifying selection as well as positive selection shapes the landscape of SCNAs during somatic evolution of cancer cells.**

Somatic copy-number alterations (SCNAs) are among the most common genomic alterations observed in cancer, and recurrent alterations have been successfully used to implicate cancer-causing genes[1]. Effectively finding cancer-causing genes using a genome-wide approach relies on our understanding of how new genome alterations are generated during the somatic evolution of cancer[3-6]. As such, we test the hypothesis that three-dimensional chromatin organization and spatial co-localization influences the set of somatic copy-number alterations observed in cancer (**Fig. 1A**, recently suggested by cancer genomic data in a study of prostate cancer[7]. Spatial proximity and chromosomal rearrangements are discussed more generally[8-11]). Unequivocally establishing a genome-wide connection between SCNAs and three-dimensional chromatin organization in cancer has until now been limited by our ability to characterize three-dimensional chromatin architecture, and the resolution with which we are able to observe SCNAs in cancer. Here, we ask whether the "landscape" of SCNAs across cancers[1] can be understood with respect to spatial contacts in a 3D chromatin architecture as determined by the recently developed HiC method for high-throughput chromosome conformation capture[2] or described theoretically via the fractal globule (FG) model (theoretical concepts[12,13], review[14]). Specifically, we investigate the model presented in **Figure 1A**, and test whether distant genomic loci that are brought spatially close by 3D chromatin architecture during interphase are more likely to undergo structural alterations and become end-points for amplifications or deletions observed in cancer.

Towards this end, we examine the statistical properties of SCNAs in light of spatial chromatin contacts in the context of cancer as an evolutionary process. During the somatic evolution of cancer[15,16] as in other evolutionary processes, two forces determine the accumulation of genomic changes (**Fig. 1A**): generation of new mutations and fixation of these mutations in a population. The rate at which new SCNAs are generated may vary depending upon the genetic, epigenetic, and cellular context. After an SCNA occurs, it proceeds probabilistically towards fixation or loss according to its impact upon cellular fitness. The fixation probability of an SCNA in cancer depends upon the competition between positive selection if the SCNA provides the cancer cell with a fitness advantage, and purifying (ie. negative) selection if the SCNA has a deleterious effect on the cell. The probability of observing a particular SCNA thus depends upon its rate of occurrence via mutation, and the selective advantage or disadvantage conferred by the alteration (**Fig. 1A**). Positive, neutral, and purifying selection are all evident in cancer genomes[17].

Our statistical analysis of SCNAs argues that both contact probability due to chromosomal organization at interphase and purifying selection contribute to the observed spectrum of SCNAs in cancer. From the full set of reported SCNAs across 3,131 cancer specimens in[1], we selected 39,568 intra-arm SCNAs (26,022 amplifications and 13,546 deletions) longer than a megabase for statistical analysis, excluding SCNAs which start or end in centromeres or telomeres. To establish that our results were robust to positive selection acting on cancer-associated genes, we analyzed a collection of 24,301 SCNAs (16,521 amplifications and 7,789 deletions, respectively 63% and 58% of the full set) that do not span highly-recurrent SCNA regions (regions listed in[1], see Methods). In the text, we present results for the less-recurrent SCNAs, and note that our findings are robust to the subset of chosen SCNAs. We perform our analysis by considering various models of chromosomal organization and purifying selection, which are then used to calculate the likelihood of the observed SCNA given the model. The likelihood framework can be used to discriminate between competing models and for performing permutation tests. The strong association we find between SCNAs and high-order chromosomal structure is not only consistent with the current understanding of the mechanisms of SCNA initiation[18], but provides insight into how spatial proximity may be arrived at via chromosomal architecture and the significance of chromosomal architecture for patterns of SCNAs observed at a genomic scale.

## Results:

*Patterns of three-dimensional chromatin architecture are evident in the landscape of SCNAs*

The initial motivation for our study was an observation that the length of focal SCNAs and the length of chromosomal loops (i.e. intra-chromosomal contacts) have similar distributions (**Figs. 1B and 1C**), both exhibiting $\sim 1/L$ scaling. Analysis of HiC data for human cells showed that the mean contact probability over all pairs of loci a distance $L$ apart on a chromosome goes as $P^{HiC}(L) \sim 1/L$ for a range of distances $L = 0.5$ to 7Mb[2]. This scaling for mean contact probability was shown to be consistent with a fractal globule (FG) model of chromatin architecture. Similarly, the mean probability to observe a focal SCNA of length $L$ goes approximately as $P^{SCNA}(L) \sim 1/L$ for the same range of distances $L = 0.5$ to 10 Mb as noted in [1]. Mathematically, the observation that the mean probability to observe an SCNA decays with length is quite significant. If two SCNA ends are chosen randomly within a chromosome arm, the mean probability to observe an SCNA is constant with increasing distance $L$. Positive selection, which tends to amplify oncogenes or delete tumor suppressors, again does not give rise to a distribution whose mean decreases with length. Either purifying selection or a length-dependent mutation rate is required to observe this result.

The connection between three-dimensional genomic architecture and SCNA structure goes beyond similarity of the distributions: loci that have higher probability of chromosomal contacts are also more likely to serve as SCNA end points (**Fig. 2**). To quantitatively determine the relationship between three-dimensional genomic architecture and SCNA, both data sets were converted into the same form. For each chromosome, we represent HiC data as a matrix of counts of spatial contacts between genomic locations $i$ and $j$ as determined in the GM06990 cell line using a fixed bin size of 1 Mb[2]. Similarly, we construct SCNA matrices by counting the number of amplifications or deletions that start at genomic location $i$ and end at location $j$ of the same chromosomes across the 3,131 tumors. **Figure 2** presents HiC and SCNA matrices (heatmaps) for chromosome 17. Away from centromeric and telomeric regions, which are not considered in this analysis, the SCNA heatmap appears similar to the HiC heatmap. In particular, domains enriched for 3D interactions also appear to experience frequent SCNA.

*Likelihood analysis demonstrates that observed SCNAs are fit best by fractal globular chromatin architecture, and all fits are improved when purifying selection is considered:*

To further test the role of chromosome organization for the generation of SCNA, we developed a series of statistical models of possible SCNA-generating processes, computed the likelihoods of these models, determined how well each model explained observed SCNAs, and performed model selection using their Bayesian Information Criterion (BIC) values (see Eq. 6 and 7). Considered models take into account different mechanisms of the generation of SCNA, with a mutation rate either: uniform in length (*Uniform*), derived from experimentally determined chromatin contact probabilities (*HiC*) or derived from contact probability in a fractal globular chromatin architecture (*FG*). We also consider models that account for fixation of the produced alterations due to selection (see Eq. 1). As non-recurrent SCNAs can potentially be either neutral or deleterious to cancer cells, both possibilities are considered during model selection. Deleterious effects of SCNAs on cellular fitness may arise from the disruption of genes or regulatory regions; as such, we expect longer SCNAs to be more deleterious. If we assume that the deleterious effect of an SCNA increases linearly with its length $L$, and consider the somatic evolution of cancer as a Moran process[16,19], we find that the probability of fixation decays roughly exponentially with length at a rate that reflects the strength of purifying selection (see Eq. 4). Consequently, we use the best fit for Equation 4 (shown in comparison with SCNA data in **Fig. 1B**), in combination with a given mutational model, to describe the effects of purifying selection. The following six models are considered: *Uniform*, *Uniform$^{+sel}$*, *HiC*, *HiC$^{+sel}$*, *FG*, *FG$^{+sel}$*, with no fitting parameters for models without selection and a single fitting parameter for selection, where the additional parameter is penalized via BIC.

Model selection provides two major observations (**Fig. 3**): First, among models of SCNA generation, a model that follows the chromosomal contact probability of the fractal globule ($\sim 1/L$) significantly outperforms other models. Second, every model is significantly improved when purifying selection is taken into account (p < .001 via bootstrapping), suggesting that SCNAs experience purifying selection. **Figure 3** presents log likelihood ratios of the models (with and without purifying selection) with respect to the uniform model. If models are fit on a chromosome-by-chromosome basis (**Supplementary Fig. 2**) we observe that for long chromosomes, the *FG* model fits better than purifying selection alone. We also find that the best-fit parameter describing purifying selection is proportional to chromosome length (**Supplementary Fig. 2A**). Since smaller values for the best-fit parameter correspond to stronger purifying selection, these two results suggest that short, gene-rich, chromosomes may experience greater purifying selection. However, we note that purifying selection proportional to the genomic length of an SCNA fits the data better than purifying selection proportional to the number of genes affected by an SCNA (**Supplementary Fig. 3**).

*Permutation analysis supports the connection between SCNAs and experimentally determined three-dimensional chromatin architecture*

After quantifying the similar statistical properties of the fractal globule and the landscape of SCNAs, we tested whether the megabase-level structure of chromosomal contacts observed in experimental HiC data was evident in the SCNA landscape. The test was performed using permutation analysis (**Fig. 4**). Since both the probability of observing an SCNA with a given length and intra-chromosomal contact probability in HiC depend strongly on distance $L$, we permuted SCNAs in a way that preserves this dependence but destroys the remaining fine structure. This is achieved by randomly assigning SCNA starting locations while keeping their lengths fixed. We find that HiC fits the observed SCNAs much better than it fits permuted SCNAs (**Fig. 4**, *p*<.001). Similar analysis within individual chromosomes shows that the fit is better for 18 of the 22 autosomal chromosomes, except for chromosomes 10, 11, 16, and 19, and is significantly better (*p*<.01) for nine chromosomes 1, 2, 4, 5, 7, 8, 13, 14, and 17 (**Fig. 4B**). While the observed amplification and deletions each separately fit better on average than their permuted counterparts (**Supplementary Fig. 5**), deletions fit relatively better than amplifications (*p*<.001 vs. *p*<.05).

## Discussion:

Our genome-wide analysis of HiC measurements and cancer SCNA finds multiple logical connections between higher-order genome architecture and re-arrangements in cancer. Using an incisive likelihood-based BIC framework, we found that: (1) probability of a 3D contact between two loci based on the FG model explains the length distribution of SCNA better than other mechanistic models or than a model of purifying selection alone; (2) comparisons with permuted data demonstrate the significant connection between megabase-level 3D chromatin structure and SCNA; (3) SCNA data reflect mutational mechanisms and purifying selection, in addition to commonly considered positive selection.

These results argue strongly for the importance of 3D chromatin organization in the formation of chromosomal alterations. While the distribution of SCNAs could conceivably depend on a complicated mutation and selection landscape, which is merely correlated with 3D genomic structure, a direct explanation via 3D genomic contacts is more parsimonious. These 3D genomic structures may vary with cell type of origin of the cancer and the specific chromatin states of these cells[20,21]; for example, re-arrangement breakpoints in prostate cancer were found to correlate with loci in specific chromatin states of prostate epithelial cells[7]. In fact, if HiC data matching the tumor cell-types of origin for the set of observed SCNAs becomes available, we may find that the cell-type specific experimental 3D contacts fit the observed distribution of SCNAs better than the fractal globule model. When we perform the permutation analysis described in **Figure 4** on SCNAs separated by cell type of origin, we find that HiC fits the observed SCNAs significantly better than it fits permuted SCNAs consistently across cancer lineages for deletions, but not for amplifications (**Supplementary Fig. 6**). Differences between amplification and deletions in model fitting, permutation testing, and across cell type of origin (**Supplementary Figs. 4, S5, S6**) may reflect differences in the strength of selection and mechanisms of genomic alteration: conceivably a simple loss of a chromosomal loop could lead to a deletion, while many amplifications may occur through more complicated processes[22] and may require interactions with homologous and non-homologous chromosomes that are not necessarily directly related to intra-chromosomal spatial proximity during interphase.

Our results suggest that a comprehensive understanding of mutational and selective forces acting on the cancer genome, not limited to positive selection of cancer-associated genes, is important for explaining the observed distribution of SCNAs. Furthermore, comparing model goodness-of-fits for the distribution of SCNAs argues that purifying selection is a common phenomenon, and that many SCNAs in cancer may be mildly deleterious "passenger mutations" (reviewed in[23,24]).

The sensitivity and relevance of comparative genomic approaches to chromosome rearrangements can only increase as additional HiC-type datasets become available. Future studies will be able to address the importance of different 3D structures to the observed chromosomal rearrangements across cell types and cell states. Perhaps even more importantly, cancer genomic sequencing data will allow for significantly more detailed analyses than the current array-based approaches, allowing for greater mechanistic insight into SCNA formation. In particular, high-throughput whole-genome sequencing data will allow for both a high-resolution analysis of interchromosomal rearrangements and yield insight into the interplay between sequence features, chromatin modifications, and 3D genomic structure.

# Methods:

*Constructing heatmaps.*

We generated SCNA heatmaps from the data of Beroukhim et al.[1] who reported a total of 75,700 amplification and 55,101 deletion events across 3,131 cancer specimens; reported events are those with inferred copy number changes >.1 or <−.1, due to experimental limitations. We restricted our analysis to intra-arm SCNAs which do not start/end near telomeric/centromeric regions separated by more than one megabase bin, giving a set of 39,568 SCNAs (26,022 amplifications and 13,546 deletions). We note that SCNAs starting/ending in centromeres/telomeres (which include full-arm gain/loss) display a very different pattern of occurrence from other focal SCNAs, particularly in terms of their length distribution, which may indicate a different mutational mechanism. Requiring a separation of greater than one megabase bin is due to resolution limits of both SCNA and HiC data (see **Supplementary Fig. 1** for details). SCNA matrices are constructed by counting the number of amplifications or deletions starting at Mb *i* and ending at Mb *j* of the same chromosomes. Similarly, HiC heatmaps were generated by counting the number of reported interactions[2] between Mb *i* and *j* of the same chromosome in human cell line GM06690.

*Mutational and Evolutionary Models of SCNA.*

To test the respective contributions of mutational and selective forces on the distribution of SCNAs, we consider the probability of observing at SCNA that starts and end at *i* and *j*

$$P_{ij} = \mu_{ij} \cdot \pi(L) \tag{1}$$

as the product of the probability of a mutation, i.e. an SCNA to occur in a single cell $\mu_{ij}$, and the probability to have this mutation fixed the population of cancer cells $\pi(L)$, where $L = |i - j|$ is the SCNA length. The mutation probability $\mu_{ij}$ depends on the model that describes the process leading to chromosomal alterations: (*Uniform*) two ends of an alteration are drawn randomly from the same chromosomal arm, giving $\mu_{ij}^{Uniform} = const$; (*HiC*) the probability of an alteration depends on the probability of a 3D contact between the ends as given by HiC data; (*FG*) the probability of alteration depends upon the probability of 3D contact according to the fractal globule model, i.e. on SCNA length *L*: $\mu_{ij}^{FG} = \mu^{FG}(L) \sim 1/L$. The probability of fixation depends on the fitness of a mutated cell as compared to non-mutated cells (see below). Each mutational model is considered by itself and in combination with purifying selection, giving six models: *Uniform*, *HiC*, *FG*, *Uniform$^{+sel}$*, *HiC$^{+sel}$*, and *FG$^{+sel}$*. For example, $P_{ij}^{FG} = \mu^{FG}(L)$, and $P_{ij}^{FG+SEL} = \mu^{FG}(L) \cdot \pi(L)$. Additional parameters are accounted for using BIC (described below).

*Effects of Selection on the Probability of Fixation*

Two major selective forces act on SCNAs: positive selection on SCNAs that amplify an oncogene or delete a tumor suppressor, and purifying selection that acts on all alterations. Purifying selection results from the deleterious effects of an SCNA that deletes or amplifies genes and regulatory regions of the genome that are not related to tumor progression. We assume that deleterious effect of an SCNA, and the resulting reduction in cells fitness $\Delta F$, is proportional to SCNA length: $|\Delta F| \propto L$.

The probability of fixation is calculated using the Moran process as model of cancer evolution[16,19]:

$$\pi(\Delta F) = \frac{1 - 1/(1+\Delta F)}{1 - 1/(1+\Delta F)^N}, \tag{2}$$

where $\Delta F$ is a relative fitness difference (selection coefficient), $N$ is the effective population size. For weakly deleterious mutations ($\Delta F < 0$, $N|\Delta F| \gg 1$, $|\Delta F| \ll 1$)

$$\pi(\Delta F) \approx \frac{\Delta F}{1 - \exp(-\Delta F N)} \tag{3}$$

Note that for many deleterious mutations this leads to an exponentially suppressed probability of fixation: $\pi(\Delta F) \propto \exp(-\Delta F N)$, a useful intuitive notion. Using a deleterious effect linear in SCNA length, $\Delta F = -L/\lambda$, we obtain the probability of fixation for purifying selection acting on an SCNA

$$\pi(L) = C \frac{L}{\exp(L/\alpha) - 1} \tag{4}$$

where C is an arbitrary constant obtained from normalization of $P(L)$, and $\alpha = \lambda/N$ is a fitting parameter which quantifies the strength of purifying selection. For gene-based purifying selection, $L$ is simply replaced by the number of genes altered. Mutations that are selectively neutral have no length dependence, so $\pi(L) = C$, and thus $P_{ij} \sim \mu_{ij}$.

*Controlling for Positive Selection*

Positive selection acting on cancer-associated genes (eg. oncogenes and tumor suppressors) presents a possible confounding factor to our analysis. To establish that our results were robust to positive selection acting on cancer-associated genes, we analyzed the subset of the 39,568 SCNAs (26,022 amplifications and 13,546 deletions) that do not span highly-recurrent SCNA regions identified by GISTIC with a false-discovery rate q-value for alteration of <.25 as listed in Beroukhim et al.[1], a collection of 24,310 SCNAs (16,521 amplifications and 7,789 deletions, respectively 63% and 58% of the full set). After SCNAs spanning highly-recurrent regions are removed, permutations are performed under the constraint that permuted SCNAs do not cross any of the highly-recurrent regions. Positive selection can also be somewhat controlled for by setting a threshold on the inferred change in copy number, to filter SCNAs that may have experienced strong positive selection in individual cancers. We note that our findings are robust to the subset of chosen SCNAs, most likely because there are many fewer driver SCNAs than passenger SCNAs (**Supplementary Fig. 7**).

*Model Selection using Poisson Loglikelihood, Bayesian Information Criterion*

Since the occurrence of a particular SCNA starting at $i$ and ending at $j$ is a rare event, we evaluate the relative ability of a model to predict the observed distribution of SCNA by calculating the Poisson Loglikelihood of the data given the model:

$$\log L(\text{SCNA} | \text{Model}) = \sum_{(i-j)>1} -P_{ij}^{\text{Model}} + SCNA_{ij} \log(P_{ij}^{\text{Model}}) \tag{6}$$

where $P_{ij}^{\text{Model}}$ is dictated by the model as explained above, and $SCNA_{ij}$ is the number of SCNAs that start and end at i and j. Since recurrent regions of amplification and deletion are different, we calculate the loglikelihood separately for amplifications and deletions, and then aggregate across these two classes of SCNAs. After the loglikelihood is calculated, models are ranked and model selection is performed using Bayesian Information Criterion (BIC). BIC penalizes models based upon their complexity, namely their number of parameters. Penalizing *k* additional parameters for *n* observed SCNAs using Bayesian Information Criterion (BIC) is straightforward:

$$BIC = \log L(\text{SCNA} | \text{Model}) - \frac{1}{2} k \log(n) \qquad (7)$$

where models with higher BIC are preferred[25]. For the permutation analysis, loglikelihood is calculated in the same way, first for the observed SCNAs, and then for permuted sets of SCNAs.

**Acknowledgements:** We thank members of the Mirny Lab for helpful conversations, in particular with Christopher McFarland regarding purifying selection and Maxim Imakaev regarding fractal globules. We thank Vineeta Agarwala, Jesse Engrietz, and Rachel McCord for helpful comments and suggestions. GF and LM are members of the NCI Physical Sciences in Oncology Center at MIT (NIH U54-CA143874).

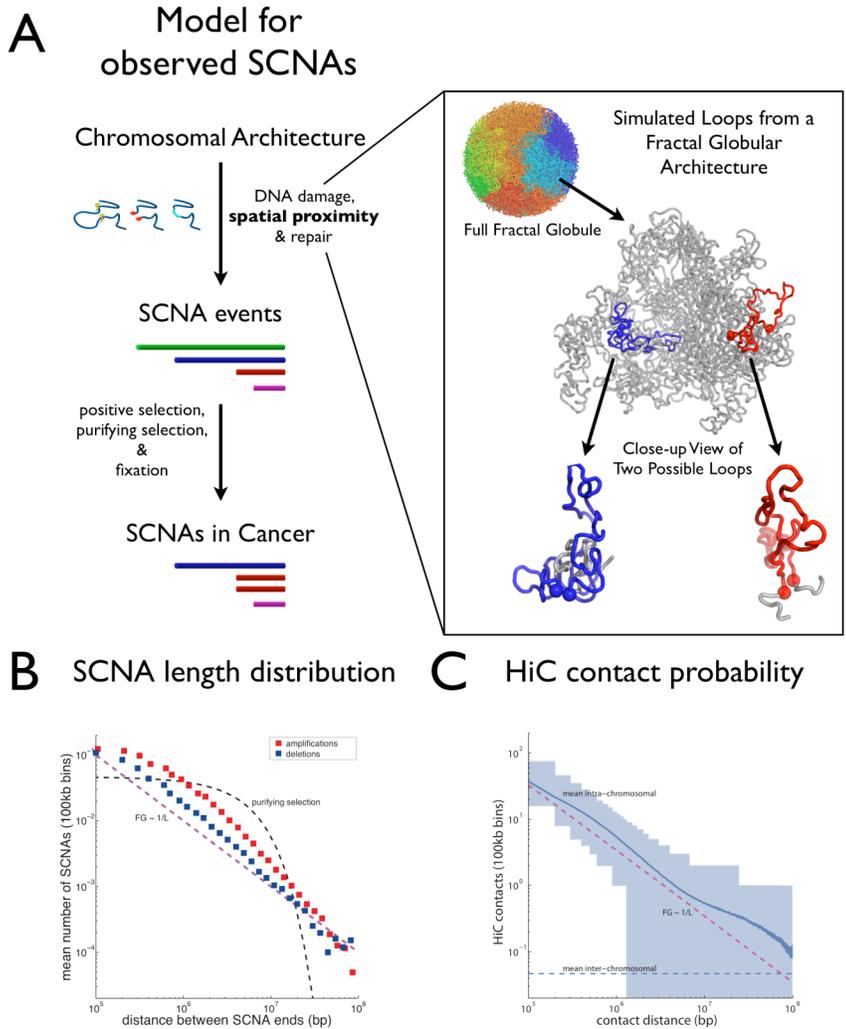

**Figure 1. 3D proximity as mechanism for SCNA formation.**
**A:** Model of how chromosomal architecture and selection can influence observed patterns of somatic copy-number alterations (SCNAs). First spatial proximity of the loop ends makes an SCNA more likely to occur after DNA damage and repair. Next, forces of positive selection and purifying selection act on SCNAs which have arisen, leading to their ultimate fixation or loss. Observed SCNAs in cancer thus reflect both mutational and selective forces. Inset illustrates looping in a simulated fractal globule architecture (coordinates from M. Imakaev). Two contact points are highlighted by spheres and represent potential end-points of SCNAs.
**B.** SCNA length distribution for 60,580 less-recurrent SCNAs (39,071 amplifications, 21,509 deletions) mapped in 3,131 cancer specimens from 26 histological types[1]. Squares show mean number of amplification (red) or deletion (blue) SCNAs after binning at 100 kb resolution (and then averaged over logarithmic intervals). Magenta dashed line shows a ~$1/L$ distribution. Dashed black line shows the best fit for purifying selection with a uniform mutation rate.
**C:** Probability of a contact between two loci distance $L$ apart on a chromosome at 100 kb resolution. The probability is obtained from intra-chromosomal interactions of 22 human chromosomes characterized by the HiC method (human cell line GM06690)[2]. Shaded area shows range from 5th and 95th percentiles for number of counts in a 100kb bin at a given distance. The mean contact probability is shown by blue line. Magenta dashed line shows ~$1/L$ scaling also observed in the fractal globule model of chromatin architecture. Blue dashed line provides a baseline for contact frequency obtained as inter-chromosomal contacts in the same dataset.

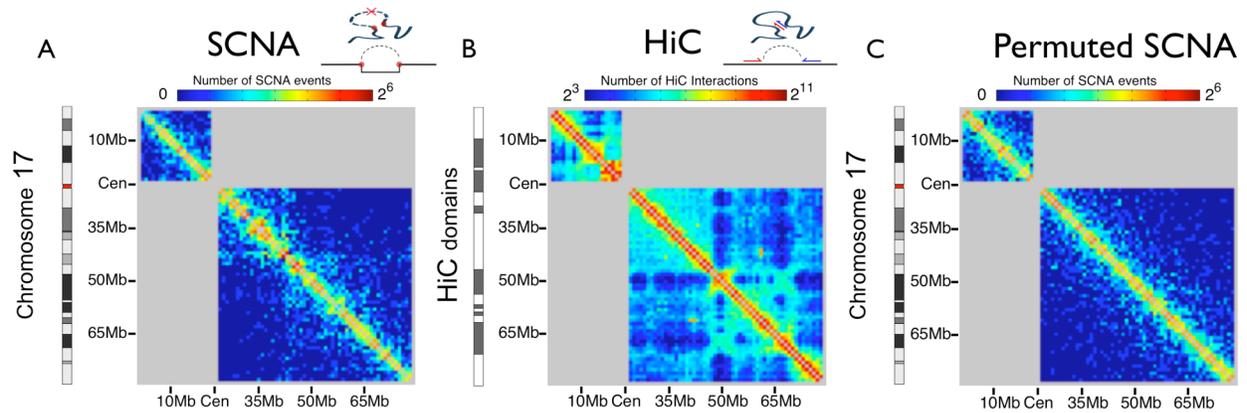

**Figure 2. Heatmaps for chromosome 17 at 1 Mb resolution**.
**A.** SCNA heatmap: the value for site ($i,j$) is the number of SCNAs starting at genomic location $i$ and ending at location $j$ on the same chromosome. Chromosome band structure from UCSC browser shown on the left side with centromeric bands in red.
**B.** HiC heatmap: site ($i,j$) has the number of reported interactions between genomic locations $i$ and $j$ at Mb resolution. HiC domain structure is shown on the left side. Domains were determined by thresholding the HiC eigenvector (as in [2], white represents open domains, dark gray represents closed domains).
**C.** Permuted SCNA heatmap: as in **A**, but after randomly permuting SCNA locations while keeping SCNA lengths fixed.
Visually, the true SCNA heatmap is similar to HiC, displaying a "domain" style organization. Cartoons above the heatmaps illustrate how mapped HiC fragments and SCNA end-points can be converted into interactions between genomic locations $i$ and $j$. Since inter-arm SCNAs, SCNAs with end-points near centromeres or telomeres, and SCNAs < 1Mb were not considered in our statistical analysis, these areas of the heatmaps are grayed out.

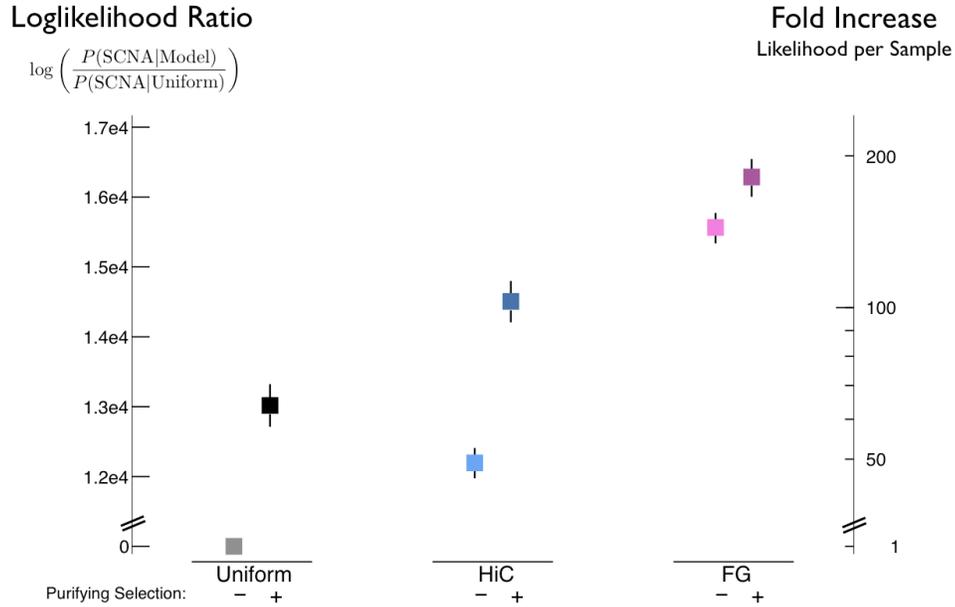

**Figure 3. Selecting a model of SCNA formation.** For each model, BIC-corrected log-likelihood is shown for the 24,310 observed SCNAs that do not span highly-recurrent SCNA regions listed in [1]. The following six models are considered: *Uniform*, *Uniform$^{+sel}$*, *HiC*, *HiC$^{+sel}$*, *FG*, *FG$^{+sel}$*. *HiC* model assumes mutation rates proportional to experimentally measured contact probabilities, while *FG* model assumes mutation rates proportional to mean contact probability in a fractal globule architecture (~1/L). Left y-axis presents BIC-corrected log-likelihood ratio for each model vs. *Uniform* model. Each model was considered with (+) and without (-) purifying selection. Right y-axis shows the same data as a fold difference in likelihood per cancer specimen (sample) vs. *Uniform*. Error bars were obtained via bootstrapping: squares represent the median values, bar ends represent the 5th and 95th percentiles. The FG model significantly outperforms other mutational models of SCNA formation, and every model is significantly improved when purifying selection is taken into account.

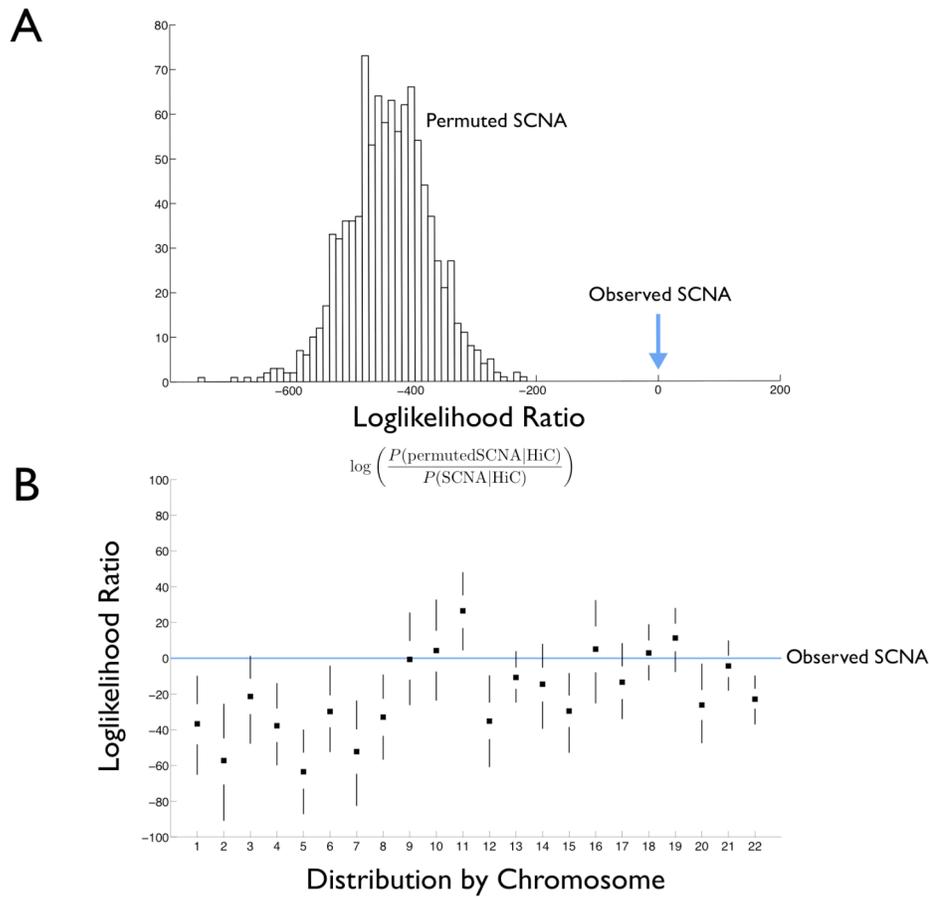

**Figure 4. Permutation analysis of the relationship between SCNAs and megabase-level structure of HiC chromosomal interactions. A.** Distribution of log-likelihood ratios for randomly permuted SCNAs given HiC vs. observed SCNAs given HiC over all 22 autosomes. Observed SCNAs (blue arrow) are fit better by HiC contact probability with *p*<.001. Permutations are performed by shuffling SCNA locations while keeping SCNA lengths fixed. **B:** Distributions of the same log-likelihood ratios for individual chromosomes (vs their corresponding observed SCNA, blue line). Squares represent median values, error bars respective represent the range from 5th to 25th percentile and 75th to 95th percentile.

# References

1. Beroukhim, R. et al. The landscape of somatic copy-number alteration across human cancers. *Nature* 463, 899-905 (2010).
2. Lieberman-Aiden, E. et al. Comprehensive mapping of long-range interactions reveals folding principles of the human genome. *Science* 326, 289-93 (2009).
3. Greenman, C. et al. Patterns of somatic mutation in human cancer genomes. *Nature* 446, 153-8 (2007).
4. Wood, L.D. et al. The genomic landscapes of human breast and colorectal cancers. *Science* 318, 1108-13 (2007).
5. Beroukhim, R. et al. Assessing the significance of chromosomal aberrations in cancer: methodology and application to glioma. *Proceedings of the National Academy of Sciences of the United States of America* 104, 20007-12 (2007).
6. The Cancer Genome Atlas Research Network. Comprehensive genomic characterization defines human glioblastoma genes and core pathways. *Nature* 455, 1061-8 (2008).
7. Berger, M.F. et al. The genomic complexity of primary human prostate cancer. *Nature* 470, 214-20 (2011).
8. Wijchers, P.J. & de Laat, W. Genome organization influences partner selection for chromosomal rearrangements. *Trends in genetics : TIG* 27, 63-71 (2011).
9. Meaburn, K.J., Misteli, T. & Soutoglou, E. Spatial genome organization in the formation of chromosomal translocations. *Seminars in cancer biology* 17, 80-90 (2007).
10. Nikiforova, M.N. et al. Proximity of chromosomal loci that participate in radiation-induced rearrangements in human cells. *Science* 290, 138-41 (2000).
11. Branco, M.R. & Pombo, A. Intermingling of chromosome territories in interphase suggests role in translocations and transcription-dependent associations. *PLoS biology* 4, e138 (2006).
12. Grosberg, A.Y., Nechaev, S.K. & Shakhnovich, E.I. The Role of Topological Constraints in the Kinetics of Collapse of Macromolecules. *Journal De Physique* 49, 2095-2100 (1988).
13. Grosberg, A., Rabin, Y., Havlin, S. & Neer, A. Crumpled Globule Model of the 3-Dimensional Structure of DNA. *Europhysics Letters* 23, 373-378 (1993).
14. Mirny, L.A. The fractal globule as a model of chromatin architecture in the cell. *Chromosome Research* 19, 37-51 (2011).
15. Nowell, P.C. The clonal evolution of tumor cell populations. *Science* 194, 23-8 (1976).
16. Merlo, L.M., Pepper, J.W., Reid, B.J. & Maley, C.C. Cancer as an evolutionary and ecological process. *Nature reviews. Cancer* 6, 924-35 (2006).
17. Lee, W. et al. The mutation spectrum revealed by paired genome sequences from a lung cancer patient. *Nature* 465, 473-7 (2010).
18. Lieber, M.R., Ma, Y., Pannicke, U. & Schwarz, K. Mechanism and regulation of human non-homologous DNA end-joining. *Nature reviews. Molecular cell biology* 4, 712-20 (2003).
19. Nowak, M.A., Michor, F. & Iwasa, Y. The linear process of somatic evolution. *Proceedings of the National Academy of Sciences of the United States of America* 100, 14966-9 (2003).
20. Mayer, R. et al. Common themes and cell type specific variations of higher order chromatin arrangements in the mouse. *BMC cell biology* 6, 44 (2005).
21. Roix, J.J., McQueen, P.G., Munson, P.J., Parada, L.A. & Misteli, T. Spatial proximity of translocation-prone gene loci in human lymphomas. *Nature genetics* 34, 287-91 (2003).
22. Hastings, P.J., Lupski, J.R., Rosenberg, S.M. & Ira, G. Mechanisms of change in gene copy number. *Nature reviews. Genetics* 10, 551-64 (2009).
23. Stratton, M.R., Campbell, P.J. & Futreal, P.A. The cancer genome. *Nature* 458, 719-24 (2009).
24. Haber, D.A. & Settleman, J. Cancer: drivers and passengers. *Nature* 446, 145-6 (2007).
25. Schwarz, G. Estimating the Dimension of a Model. *The Annals of Statistics* 6, 461-464 (1978).

# Supplementary Information

**Supplemental Figures**

**A. Limits of SCNA Reolution**

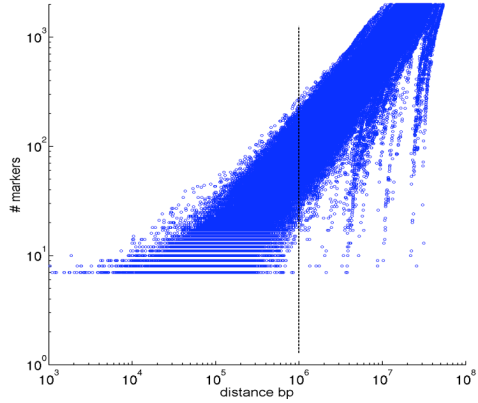

**Figure S1: Resolution of SCNA**

In [1], only SCNAs covering at least 7 probes were reported. Since SNP arrays are not evenly spaced across the genome, depending upon the region of the genome, 7 probes span anywhere from a kilobase to a megabase. Dashed line indicates that the SCNAs considered were those larger than 1Mb.

**B: Model Fits Chromosome-by-Chromosome**

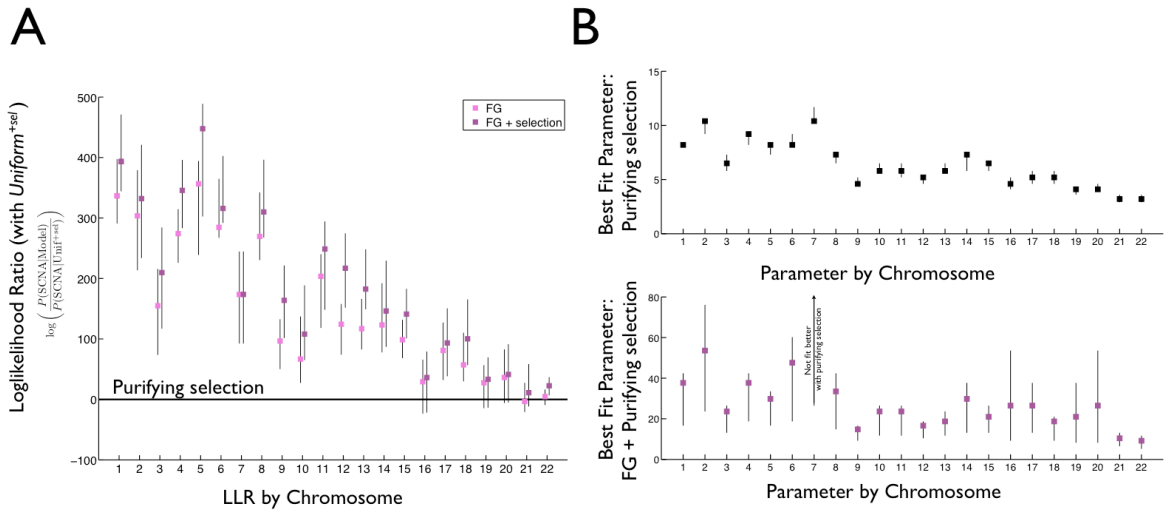

# Figure S2: fits on chromosome-by-chromosome basis

**A:** Model fitting chromosome-by-chromosome via Poisson Loglikelihood for SCNAs that do not span highly-recurrent SCNA regions listed in [1]. BIC penalized LLR for indicated models vs. model where two ends of an alteration are drawn randomly from the same chromosomal arm and experience purifying selection (*Uniform*$^{+sel}$). Error bars for the fits obtained via bootstrapping, and respectively represent the range from 5$^{th}$ to 95th percentile. For long chromosomes, *FG* alteration probability ~ 1/L clearly fits better than purifying selection, *Uniform*$^{+sel}$, on a chromosome-by-chromosome basis.

**B:** Parameter values (in Mb) for best fitting exponential distribution when purifying selection is fit on a chromosome-by-chromosome basis. Top (black squares): simple purifying selection (*Uniform*$^{+sel}$). Bottom (magenta: squares): purifying selection and mutation rate from fractal globule (*FG* $^{+sel}$). Error bars obtained via bootstrapping. Parameter values for purifying selection (best fitting exponential distribution) are proportional to chromosome length for both simple purifying selection, *Uniform*$^{+sel}$ and *FG* $^{+sel}$. This is also true *HiC* $^{+sel}$. As smaller paramter values indicate stronger purifying selection, the relationship between parameter values and chromosome length suggests that shorter, more gene rich chromosomes, may experience greater purifying selection

**C: Gene-Based Purifying Selection & Cancer Associated Genes**

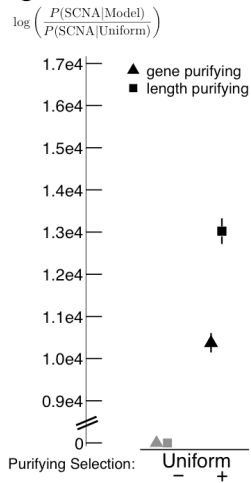

**Figure S3**. For each model, BIC-corrected loglikelihood for the set of non-recurrent SCNAs is shown. Comparison of gene-based purifying selection (triangles) with length-based purifying selection (squares). Each model is considered with and without purifying selection. Thus *HiC* w/o (-) purifying selection is the same for gene-based or length-based purifying selection. Y-axis shows loglikelihood Ratios for SCNA data vs. *Uniform*. Across model types, the best-fit parameter for length-dependent purifying selection fits the data better than purifying selection acting on the number of genes affected by a SCNA. Error bars obtained via bootstrapping: symbol represents the median, bar ends represent the 5th and 95th percentiles.

**D: Amplifications vs. Deletions**

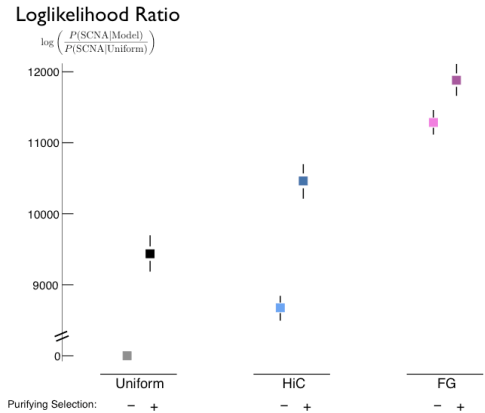 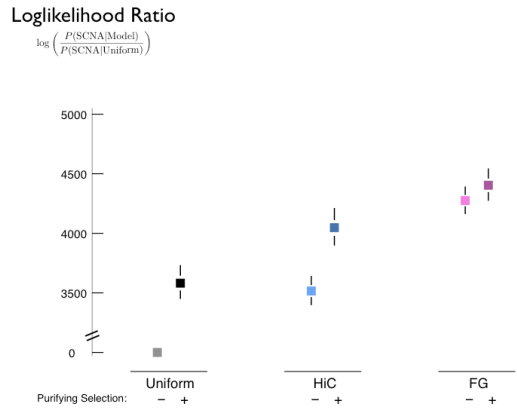

**Figure S4.** Poisson Loglikelihood Ratios for SCNA data.

**A:** Amplifications: model loglikelihood for 16,521 observed amplification SCNAs in the non-recurrent set vs. *Uniform*.

**B:** Deletions: model loglikelihood for 7,789 observed deletion SCNAs in the non-recurrent set vs. *Uniform*

The following six models are considered: *Uniform*, *Uniform$^{+sel}$*, *HiC*, *HiC$^{+sel}$*, *FG*, *FG$^{+sel}$*. *HiC* model assumes mutation rates proportional to experimentally measured contact probabilities, while *FG* model assumes mutation rates proportional to contact probability in a fractal globule architecture (~1/L). Left y-axis presents BIC-corrected loglikelihood ratio for each model vs. *Uniform* model. Each model was considered with (+) and without (-) purifying selection. Error bars were obtained via bootstrapping: square represents the median, bar ends represent the 5th and 95th percentiles. The FG model significantly outperforms other mutational models of SCNA formation for amplifications and deletions. However, *Uniform$^{+sel}$* does not outperform *HiC* for deletions and *FG$^{+sel}$* does not fit significantly better than *FG* for deletions. We note that since there are more amplifications than deletions, the aggregate likelihood ratios vs. the *Uniform* model are greater for amplifications. The relatively poorer performance of *HiC* for amplifications may reflect additional selective or mutational pressures acting on amplifications vs. deletions. Variable mutation rates at the megabase-scale could obscure a relationship with megabase details of chromatin structure.

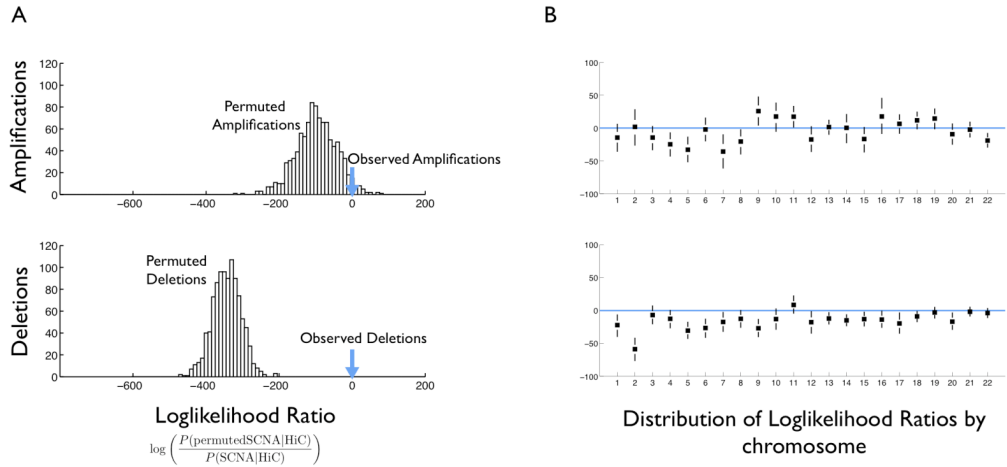

**Figure S5**: **Permutation tests for amplifications and deletions.**
**A:** distribution of loglikelihood ratios for randomly permuted SCNAs given HiC vs. observed SCNAs across all chromosomes, separated into results for the 16,521 amplifications (top) and 7,789 deletions (bottom). Observed amplifications fit better by HiC contact probability with $p<.05$, observed deletions are fit better with $p<.001$.
**B:** Distributions of same loglikelihood ratios for individual chromosomes (22 autosomes) vs. observed SCNAs (blue line). Squares represent median values, error bars respective represent the range from 5th to 25th percentile and 75th to 95th percentile. On average, the probability of the observed deletions given *HiC* is higher than permuted deletions for each chromosome except chromosome 11.

**E: Permutations by cell-type (cancer lineage)**

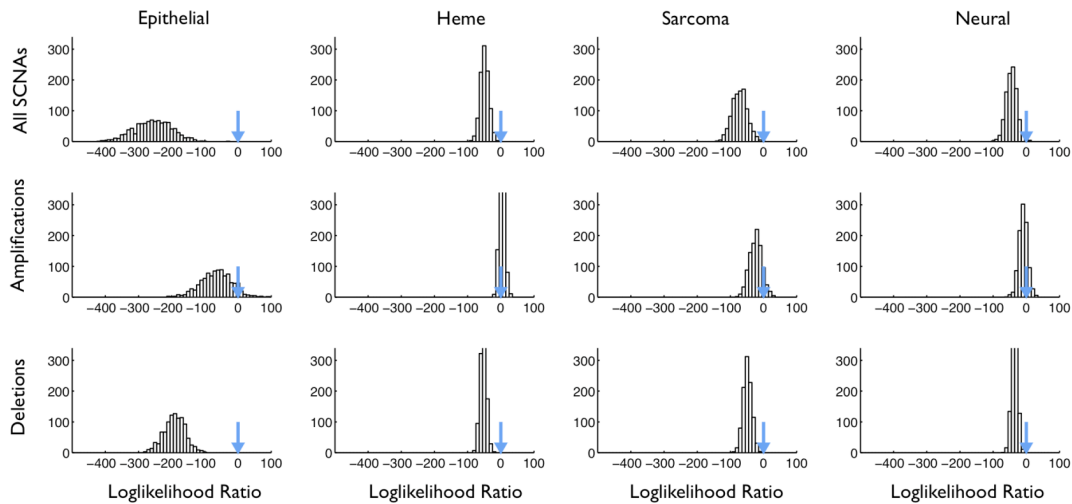

**Figure S6**: **Permutation tests by cancer lineage (cell type of origin)**
Distribution of loglikelihood ratios for randomly permuted SCNAs given HiC vs. observed SCNAs across all chromosomes, for all SCNAs (top), amplifications (middle) and deletions (bottom). Cancers are separated into epithelial, heme, sarcoma, and neural lineages as indicated in Supplemental Figure 7 in [1]. Deletions are significant across all cancer subtypes.

**F: Results are robust to choice of SCNAs**

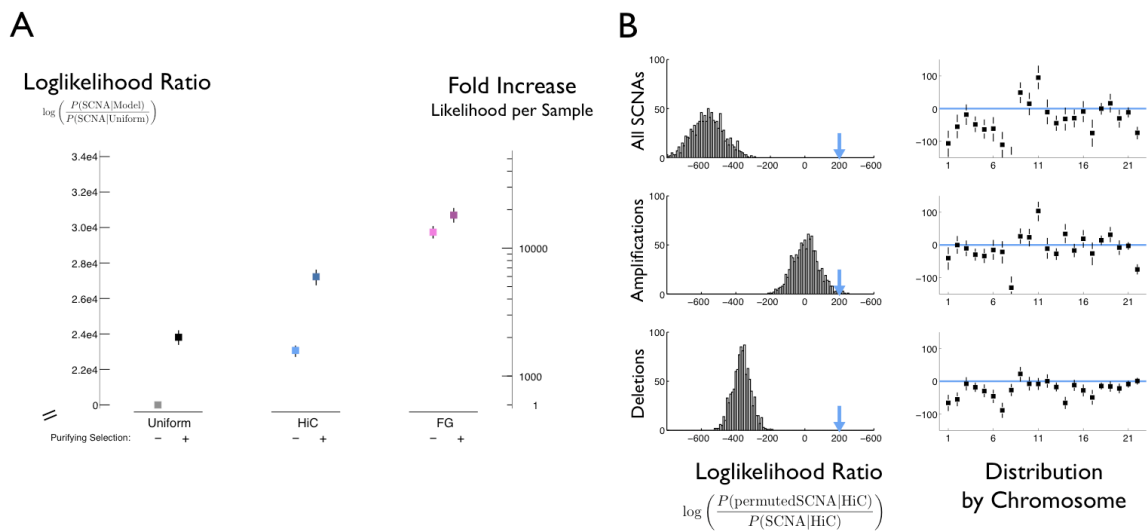

**Figure S7. Model selection and permutation analysis are robust to choice of SCNAs**
**A:** Same as **Figure 3**, but including SCNAs which span significant GISTIC peaks of recurrent SCNA (39,568 SCNAs, 26,022 amplifications and 13,546 deletions). Left y-axis presents BIC-corrected loglikelihood ratio for each model vs. *Uniform* model. Each model was considered with (+) and without (-) purifying selection. Right y-axis shows the same data as a fold difference in likelihood per cancer specimen (sample) vs. *Uniform*. Error bars were obtained via bootstrapping: square represents the median, bar ends represent the 5th and 95th percentiles. The FG model significantly outperforms other mutational models of SCNA formation, and every model is significantly improved when purifying selection is taken into account.
**B:** (top row) same as **Figure 4**, but including SCNAs that span significant GISTIC peaks of recurrent SCNA. (middle and bottom row) same as **Figure S5**. Left column shows distribution of loglikelihood ratios for randomly permuted SCNAs given HiC vs. observed SCNAs given HiC aggregated over all 22 autosomes. Observed SCNAs are indicated with a blue arrow. Right column shows the distributions for individual chromosomes. Squares represent median values, error bars respective represent the range from 5th to 25th percentile and 75th to 95th percentile.

1. Beroukhim, R. et al. The landscape of somatic copy-number alteration across human cancers. *Nature* 463, 899-905 (2010).
2. Lieberman-Aiden, E. et al. Comprehensive mapping of long-range interactions reveals folding principles of the human genome. *Science* 326, 289-93 (2009).